\documentstyle[aps,multicol,psfig]{revtex}

\def\be{\begin{equation}}
\def\ee{\end{equation}}
\def\bea{\begin{eqnarray}}
\def\eea{\end{eqnarray}}
\def\bean{\begin{eqnarray*}}
\def\eean{\end{eqnarray*}}
\def\half{\frac{1}{2}}

\def\bra{\langle}
\def\ket{\rangle}

\newcommand{\al}{\alpha}

\newcommand{\gm}{\gamma}

\newcommand{\lm}{\lambda}

\newcommand{\sg}{\sigma}

\newcommand{\dslash}{/\!\!\! \partial}

\newcommand{\Aslash}{/\!\!\!\! A}

\begin{document}

\title{
Non-Equilibrium Dynamics with Fermions on a Lattice in Space 
and Time\footnote{Poster presented at the 5th International 
Workshop on Thermal Field Theories and Their Applications, Regensburg, 
Germany, August 10-14, 1998.}}

\author{Gert Aarts$^1$ and Jan Smit$^{1,2}$}

\address{
$^1$ Institute for Theoretical Physics, Utrecht University, 
Princetonplein 5, 3508 TA Utrecht, the Netherlands\\
$^2$ Institute for Theoretical Physics, 
University of Amsterdam, 
Valckenierstraat 65, 1018 XE Amsterdam, the Netherlands\\
{\rm aarts@phys.uu.nl, jsmit@phys.uva.nl}
}

\date{September 1, 1998}

\maketitle

\begin{abstract}
We consider the dynamics of the $1+1$ dimensional abelian Higgs model 
with axially coupled fermions, in the large $N_f$ limit, on a lattice in 
space and real-time. We allow for inhomogeneous classical Bose fields. In 
order to deal with the lattice doublers, we use Wilson's lattice fermions. 
The lattice formulation leads to a stable integration algorithm.
We illustrate the implementation with numerical results.
\end{abstract}
  
\begin{multicols}{2}
\section{Introduction}

Real-time dynamics of quantum fields plays an 
important role in the early universe, in heavy ion collisions and in 
condensed matter.
For example, a detailed understanding of the rate of sphaleron 
transitions at high temperature is relevant for baryogenesis. This is a 
non-perturbative problem, which should 
be treated preferably directly in real-time.

However, as is usually the case, in order to do actual 
calculations, approximations have to be made. 
The classical approximation \cite{GrRu88,abh,TaSm98,su2,AaSm97} is based on 
the observation that  
at sufficiently high temperature the relevant long distance modes in the 
quantum theory behave classically. Hence it makes sense to study the 
classical equations of motion. The advantage is that it is 
(almost) straightforward to implement it numerically, and that e.g.  
sphaleron transitions are easy to find.
However, it has become clear that Hard Thermal Loop (HTL) effects play an 
important role, and that it is necessary to incorporate them 
in a classical field and/or classical particle approach 
\cite{classhtl1,classhtl2}.

Other possibilities are large $N$ or Hartree approximations
\cite{CoMo1,CoMo2,largeN,baacke}, which have been applied to e.g.
(inflationary) scalar models and gauge theories. In this formulation 
the fields are split in a classical part and a quantum field, and they 
are coupled via semiclassical equations of motion, which are exact in the 
limit $N\to \infty$. The advantage is that quantum effects, including the  
backreaction on the classical fields, are taken into account. However, in 
actual numerical calculations, the emphasis has been on the case of 
homogeneous classical fields \cite{largeN,baacke}. 
This excludes inhomogeneous classical field configurations, such as the 
sphaleron.

Since we want to improve both the classical approximation by including 
quantum effects (such as HTL's in $3+1$ dimensions) and the large $N$ 
approach by 
relaxing the classical field being homogeneous, we consider here the second 
approach with inhomogeneous classical fields.
 
After a discussion of the model, we describe the effective equations 
of motion. They are defined on a lattice in space and time, to have gauge 
invariance and numerical stability. We illustrate this with some 
numerical results.

\section{Abelian Higgs model with axially coupled fermions}

We consider the abelian Higgs model in $1+1$ dimensions, with axially 
coupled fermions. The (continuum) action is 
\bean
S = -\int d^2 x \Big[\frac{1}{4e^2}F_{\mu\nu}F^{\mu\nu} 
   + |D_\mu \phi|^2 + 
\lambda(|\phi_x|^2 - \half v^2)^2
&&\\
 + \bar\psi(\dslash + \half i\Aslash\gm_5)\psi + G
\bar\psi(\phi^*P_L+\phi P_R)\psi\Big].
&&
\eean
It has a number of properties in common with the 
Standard Model.
The bosonic vacuum is non-trivial, and is labeled with the Chern-Simons 
number $C=-\int dx^1 A_1/2\pi$, being integer at a vacuum.
The vacua are separated by a finite energy sphaleron 
barrier, where $C$ is half integer.
There is a global vector symmetry, $\psi \to 
e^{i\xi}\psi$, which leads to a classically conserved vector current
$j^\mu_f = i\bar\psi\gm^\mu\psi$.
However, the corresponding charge, the fermion number $Q_f=\int dx^1 
j^0_f$, is not conserved in  the quantum theory, because of the anomaly: 
$Q_f(t)-Q_f(0) = -(C(t)-C(0))$. 
The bosonic part of this model has been used before to test the classical 
approximation \cite{abh,TaSm98}.

Being in $1+1$ dimensions has a few consequenses. 
The typical HTL expressions are subdominant, so that these play a less 
important role \cite{TaSm98}.
Also the divergences are less severe, only the Higgs self energy is 
divergent, and only due to Higgs and fermion one loop contributions.
And it is much easier to deal with numerically than $3+1$ dimensional models.

\section{On the lattice in real-time}

In order to formulate the model on a lattice, it is 
convenient to first put the model in a more suitable form. 
By performing charge conjugation on the right-handed 
fermions only,
$\psi_R \to (\bar\psi_R{\cal C})^T, \psi_L \to \psi_L$, where ${\cal C}$ is 
the charge conjugation matrix, we write the model 
as one with a vector gauge symmetry.
The fermionic part of the action becomes the usual Schwinger model with 
a Majorana-like Yukawa coupling
\bean S_f 
&=&
 -\int d^2x\Big[
\bar\psi (\dslash -\half i\Aslash)\psi 
\\
&&\;\;\;\;\;\;\;\;
+
\half G\psi^T{\cal C}^\dagger\phi^*\psi 
-\half G\bar\psi{\cal C}\phi\bar\psi^T\Big].
\eean
The anomalous current is now axial, $j^\mu_5 =
i\bar\psi\gm^\mu\gm_5\psi$, and  the anomaly equation reads
\[Q_5(t)-Q_5(0) = C(t)-C(0).\]
Since the Yukawa term has become Majorana-like, 
it is convenient to work with a real 4-component Majorana field $\Psi$.

It is now straightforward to put this theory on a space-time lattice with 
lattice spacing $a_0, a_1=a$, using Wilson's lattice fermions,  known from 
euclidean lattice field theory. Lattice equations of motion and currents 
associated to global symmetries follow in the usual way.

\section{Effective equations of motion}

Effective equations of motion can be derived by considering $N_f$ 
fermion fields with $N_f\to \infty$. After integrating out the fermions 
and rescaling $\phi$ and $v$ with $\sqrt{N_f}$, $e$ and $G$ with 
$1/\sqrt{N_f}$, and $\lm$ with $1/N_f$, the resulting bosonic effective 
action is proportional to $N_f$, and the remaining bosonic path integral 
can be approximated with a saddle point expansion. In leading order this 
results in the semiclassical Maxwell-like equations of motion \cite{CoMo2}.
In this approximation, the Bose fields are treated as classical fields, and 
the fermions, that are fully quantized, live in their background.
The initial conditions for the original (unscaled) $\phi$ and $eA_\mu$
have evidently to be such that they are of order $\sqrt{N_f}$.

When the classical field is constrained to be homogeneous, the quantum 
fluctuations are treated in practise with a mode function expansion 
\cite{largeN,baacke}. Since we want to 
allow also inhomogeneous Bose fields, we use a 'generalized' mode function 
expansion, i.e. at $t=0$, the fermion field is expanded in plane waves in 
order to be able to specify the initial quantum state, but  
for $t>0$ the space dependence is determined by the field equations.
The result is that the mode 
functions have both a momentum label $p$ corresponding to the 
initial state, and a general space (and time) dependence, written as 
$x=(t,x^1)$. 

We expand the field as
\[
\Psi_x = \frac{1}{\sqrt{L}}\sum_{p,\al}\left[ b_p(\al)w_p(x;\al)
+ b^\dagger_p(\al)w^*_p(x;\al)\right],
\]
with the initial condition
\[   
w_p(t=0, x^1;\al) = e^{ipx^1}w_p(\al).
\]
$w_p(\al)$ is an eigenspinor of the Dirac hamiltonian 
at $t=0$, and $\al=\{+,-\}$ labels the independent eigenspinors.
Expectation values of the time independent creation and annihilation 
operators $b_p^{(\dagger)}(\al)$ specify the quantum state at $t=0$. 
We take either a vacuum or a thermal initial state, summarized by
$\sg_p(\al) \equiv \bra [b_p(\al),b_p^\dagger(\al)]\ket
= 1-2f_p(\al)$, with $f_p$ the Fermi-Dirac distribution.

The bosonic equations are (in the temporal gauge $A_0=0$)
\bean
&&\partial'_0\partial_0 A_{1x} = e^2 (j_{h_x}^1 + \bra
j_{f_x}^1\ket),\\
&&\partial'_1\partial_0 A_{1x} = -e^2 (j_{h_x}^0 + \bra 
j_{f_x}^0\ket),\\ &&
\partial'_0\partial_0 \phi_x = D'_1 D_1\phi_x - 2\lambda
(|\phi_x|^2 - \half v^2_B)\phi_x + G\bra F_x\ket.
\eean  
The second equation is Gauss' law. The (primed) derivatives denote 
forward (backward) lattice derivatives. $j_h^\mu$ is the classical 
Higgs contribution to the current.

The fermion current $\bra j_f^\mu\ket$ and force due to the 
fermions on the scalar field $\bra F\ket$ can be written in terms of mode 
functions. The mode functions 
themselves obey the lattice Dirac equation in the presence of the Bose 
fields. This gives a closed set of equations. 

Also observables such as the fermion vector and axial charge are 
expressed in terms of the mode functions. 
As an example, we show here the axial charge density $\bra j^0_{5_x}\ket$ 
on the lattice
\bea
\nonumber
\bra j^0_{5_x}\ket &=&
-\frac{1}{4L}\sum_{p,\al} \sg_p(\al)\Big[
w_p^\dagger(t,x^1;\al)\gm_5\rho_2 w_p(t+ a_0, x^1;\al) \\
\label{eqq5}
&&\;\;\;\;\;+ 
w_p^\dagger(t+ a_0, x^1;\al)\gm_5\rho_2 w_p(t, x^1;\al)
\Big].
\eea
Here $\rho_2$ is a matrix appearing in the Majorana description.
 
The equations of motion still contain the divergence of the quantum 
theory. Since we only integrated out the fermions, and we treat the 
scalar field completely classical, only the fermion loop divergence 
enters in the equations. $\bra F_x\ket$ is a logarithmically 
divergent sum. It's divergence is canceled with the appropriate bare $v_B^2$.

\section{Numerical results} 

We have used the following initial conditions.
The initial fermion mode functions are eigenspinors of the Dirac hamiltonian 
in the presence of a vacuum configuration of Bose fields, i.e. $A_1=0, 
\phi=v_R/\sqrt{2}$, and we take a vacuum initial state for the 
creation and annihilation operators, i.e. $\sg_p(\al)=1$.
The Higgs field is initially homogeneous, $\phi=v_R/\sqrt{2}$, but 
for its time derivative, $\partial_0\phi=\pi$, only the low momentum modes 
have a non-zero amplitude. The high momentum modes of $\pi$ are initially 
zero. This makes the classical field inhomogeneous.
Both the Chern-Simon number and its time derivative have a non-zero value 
at $t=0$. 
The initial configuration is chosen in such a way that it obeys  Gauss' 
law. This completely specifies the spatial dependence of $\partial_0 A_1$.

\vspace{-0.4cm}
\begin{figure}
\psfig{figure=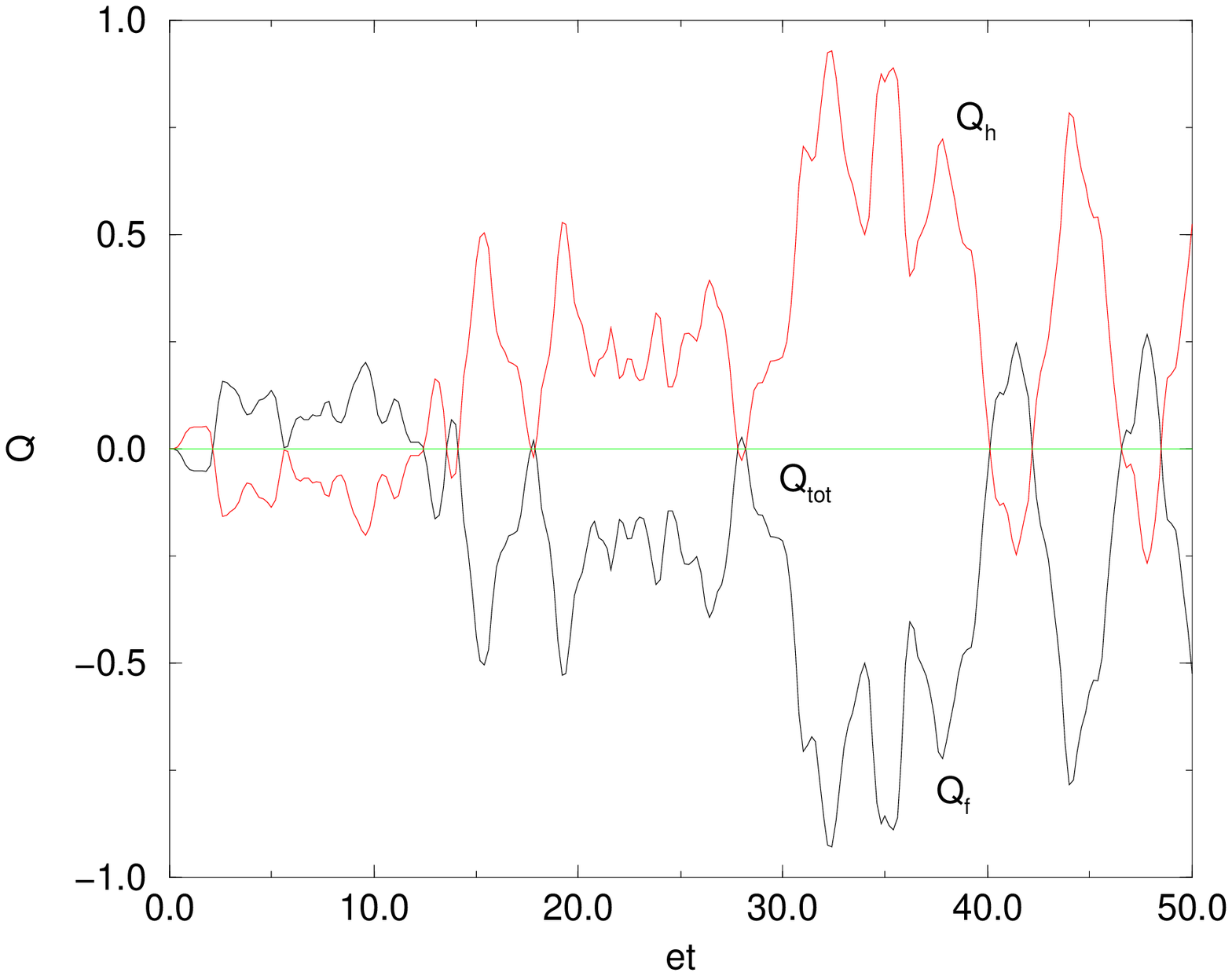,width=8.5cm}
\vspace{-0.4cm}
\caption{\narrowtext $Q_h$, $Q_f$ and $Q_{\rm tot}=Q_f+Q_h$ versus $et$.}
\label{fig1}
\psfig{figure=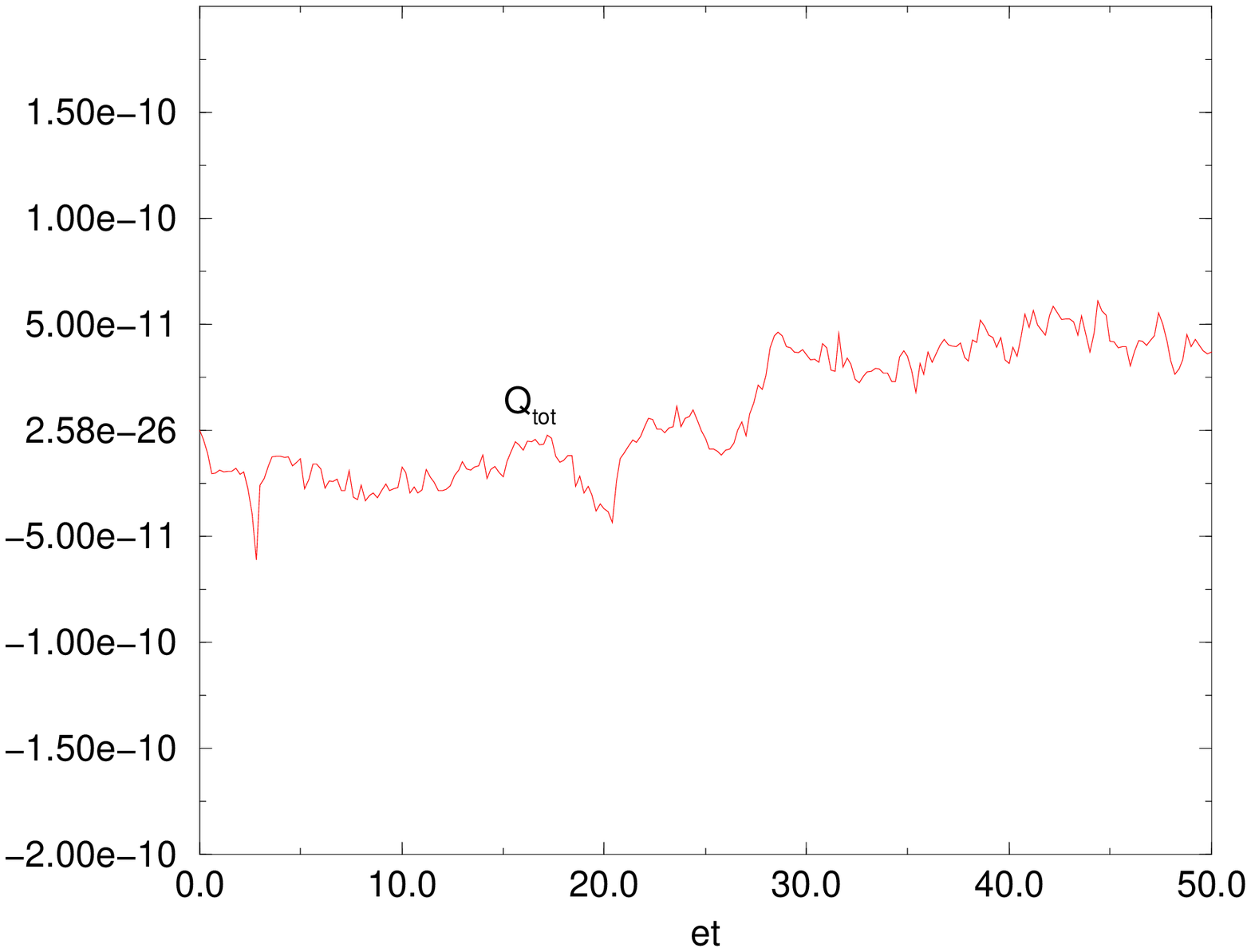,width=8.5cm}
\vspace{-0.4cm}
\caption{\narrowtext Total conserved charge $Q_{\rm tot}$ versus $et$.}
\label{fig1a}
\end{figure}

The quantum divergence in the equations is treated in the following way. 
The renormalized Higgs expectation value $v_R$ is fixed, and 
then the corresponding bare $v_B$ is determined using the equation 
for the scalar field at $t=0$. We checked that this leads to  convergent 
physical results when the lattice spacing $a$ is made smaller.

Every dimensionful parameter is written in units of $e$. The 
parameters are $\lambda/e^2=G/e=0.25$ and the 
renormalized Higgs expectation value is $v_R^2=10$. The physical size 
is $eL=3.2$ ($L=aN$). Furthermore, the temporal lattice 
spacing $a_0/a=0.01$, and the number of lattice points  $N=64$.

We now discuss the numerical results. Gauss' law demands that 
the total charge $Q_{\rm tot}(t) =Q_h(t)+Q_f(t)$ is zero. Our initial 
conditions 
are such that both the Higgs charge, $Q_h(0)$, and the fermion charge, 
$Q_f(0)$, are separately zero.
Under real-time evolution, having an inhomogeneous system, the individual 
charges do not stay zero, but  the sum vanishes (up to machine 
precision). This is demonstrated in Figs.~\ref{fig1}, \ref{fig1a}. 
Since we solve a large set of partial differential equations (in 
particular there is an equation for every mode function, and there are 
$2N$ mode functions), this is non-trivial.

\vspace{-0.4cm}
\begin{figure}
\psfig{figure=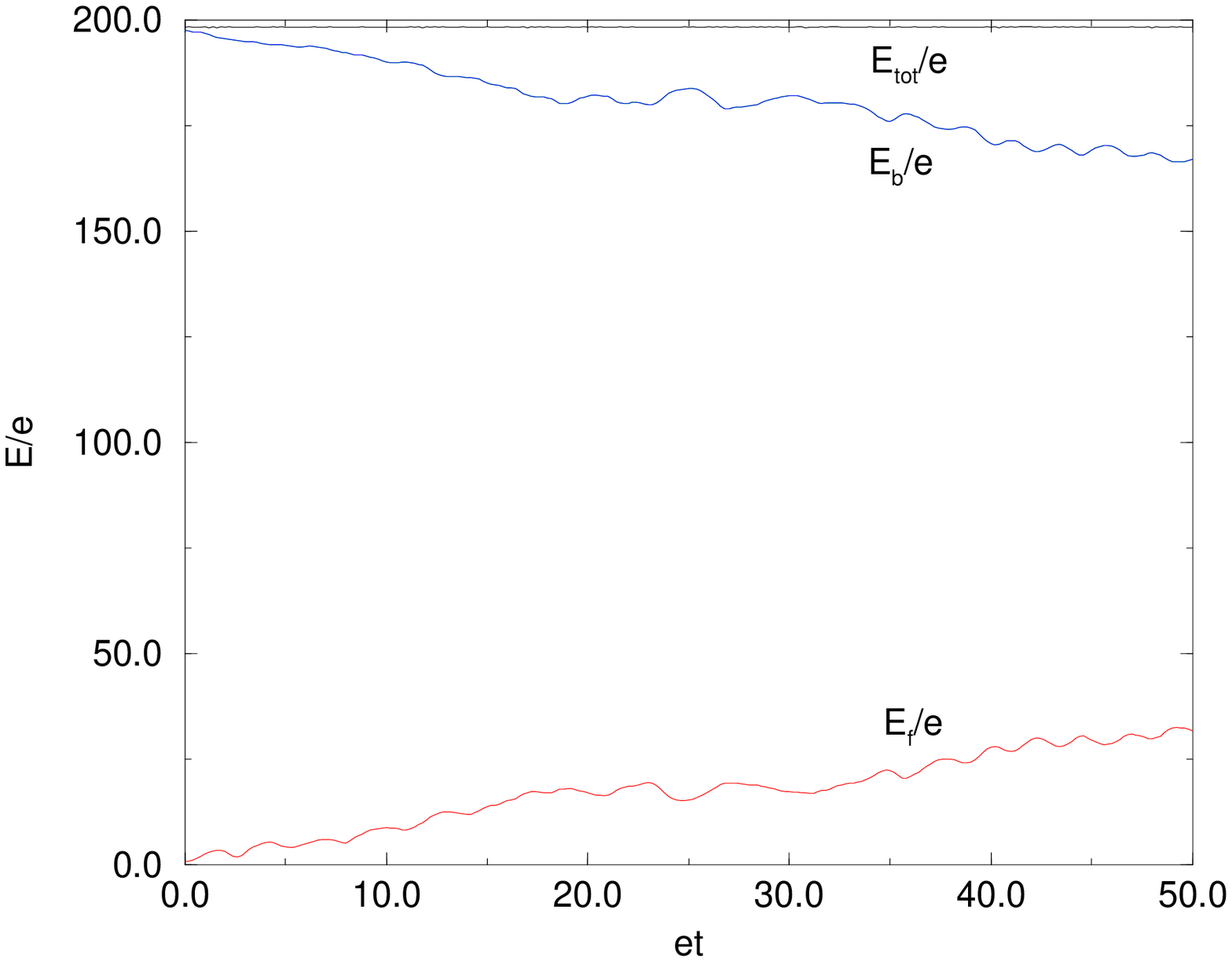,width=8.5cm}
\vspace{-0.4cm}
\caption{\narrowtext Bosonic energy, $E_b/e$, fermionic energy, $E_f/e$, and 
total energy $E_{\rm tot}/e=(E_b+E_f)/e$
versus $et$.}
\label{fig2}
\psfig{figure=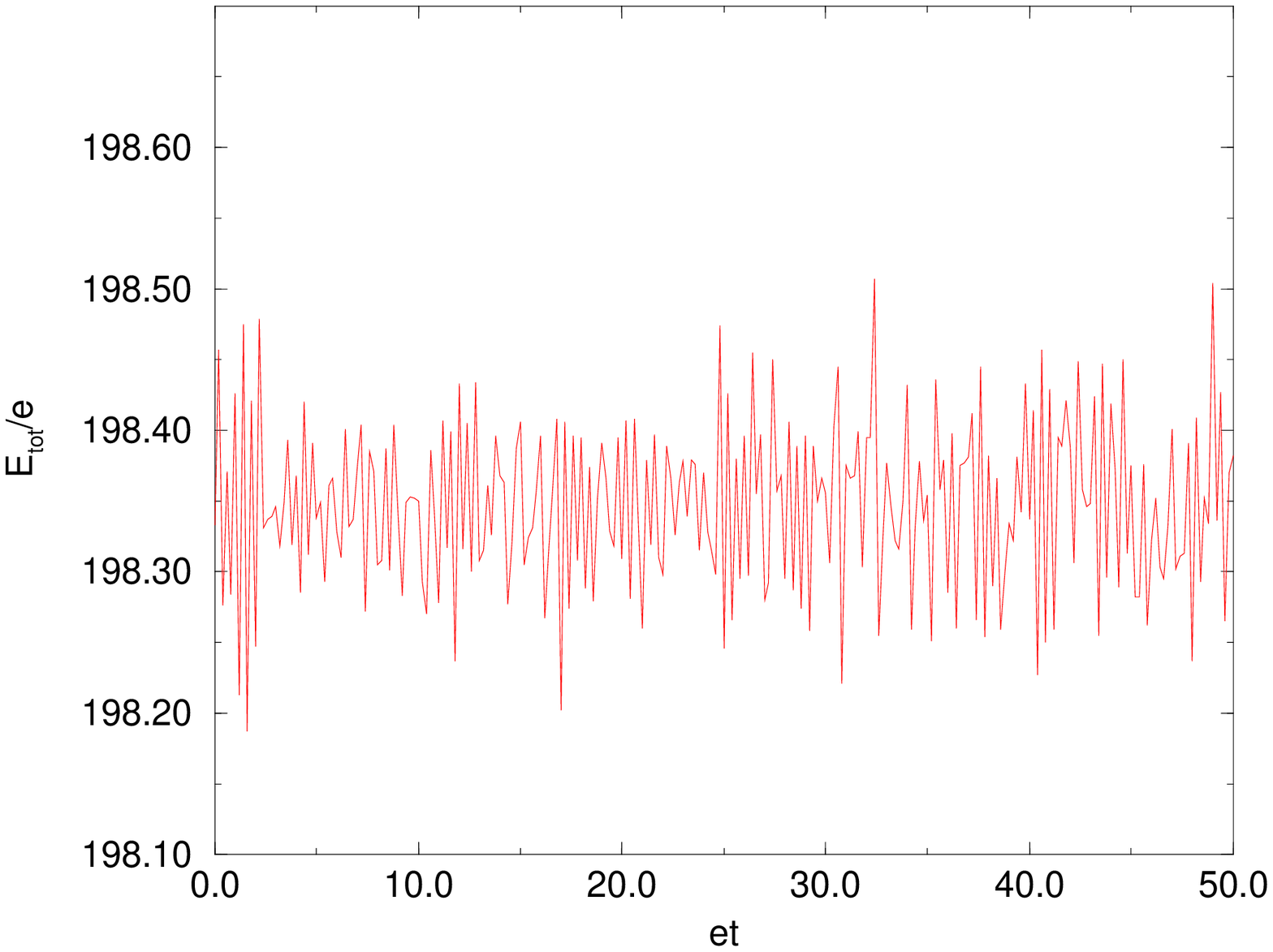,width=8.5cm}
\vspace{-0.4cm}
\caption{\narrowtext
Total conserved energy $E_{\rm tot}/e$ versus $et$.}
\label{fig2a}
\end{figure}

Also the total energy $E_{\rm tot}=E_b+E_f$ is conserved, but not to such 
 high accuracy as the charge. This is shown in Figs.~\ref{fig2}, 
\ref{fig2a}. The fermions are initialized in a vacuum state 
($E_f(0)=0$). All the initial energy is contained in the low momentum 
modes of the Bose fields. This is clearly a non-equilibrium situation, 
and energy is transferred into the fermion degrees of freedom 
during time evolution.

\begin{figure}
\psfig{figure=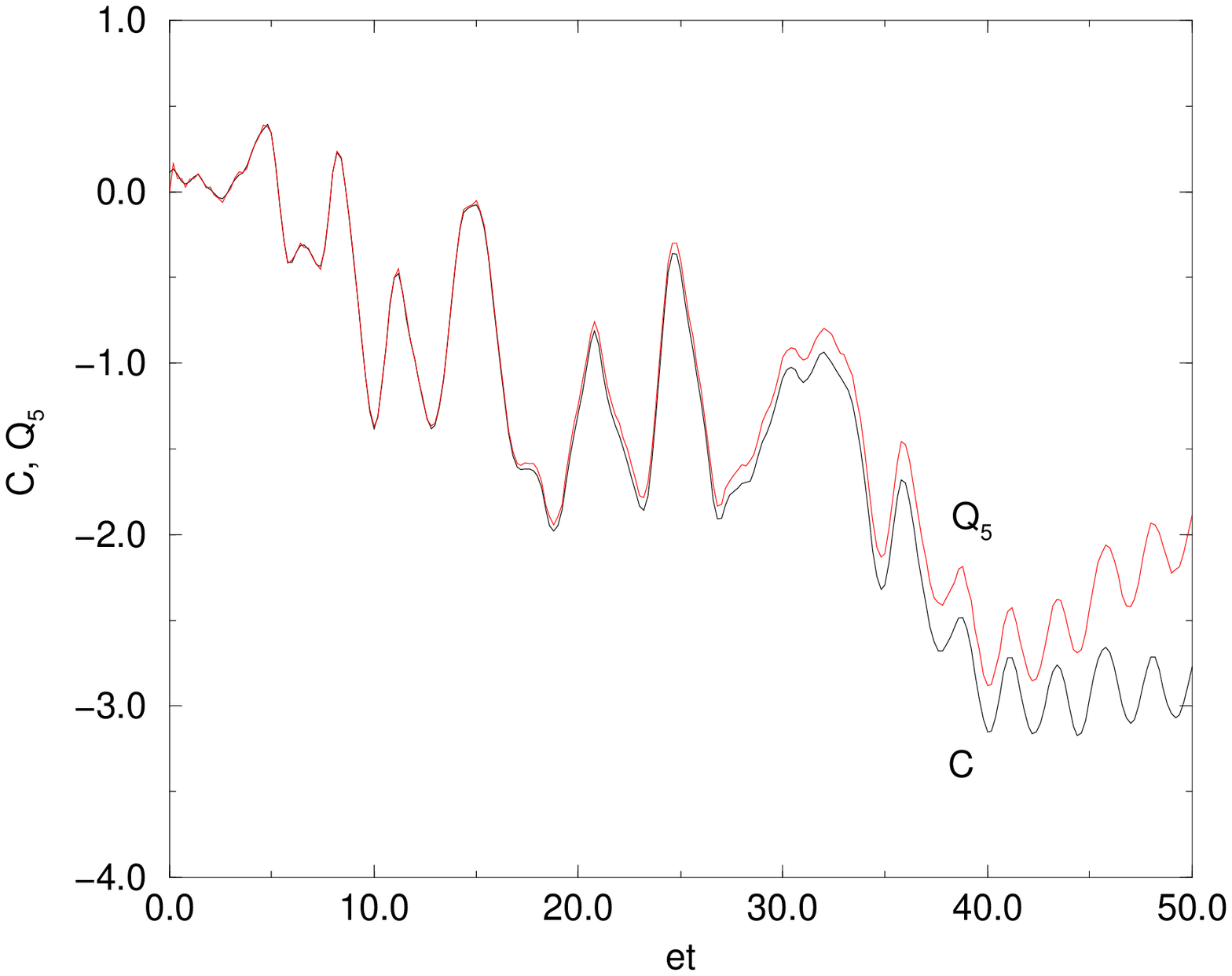,width=8.5cm}
\vspace{-0.4cm}
\caption{\narrowtext $C$ and $Q_5$ versus $et$.}
\label{fig3}
\psfig{figure=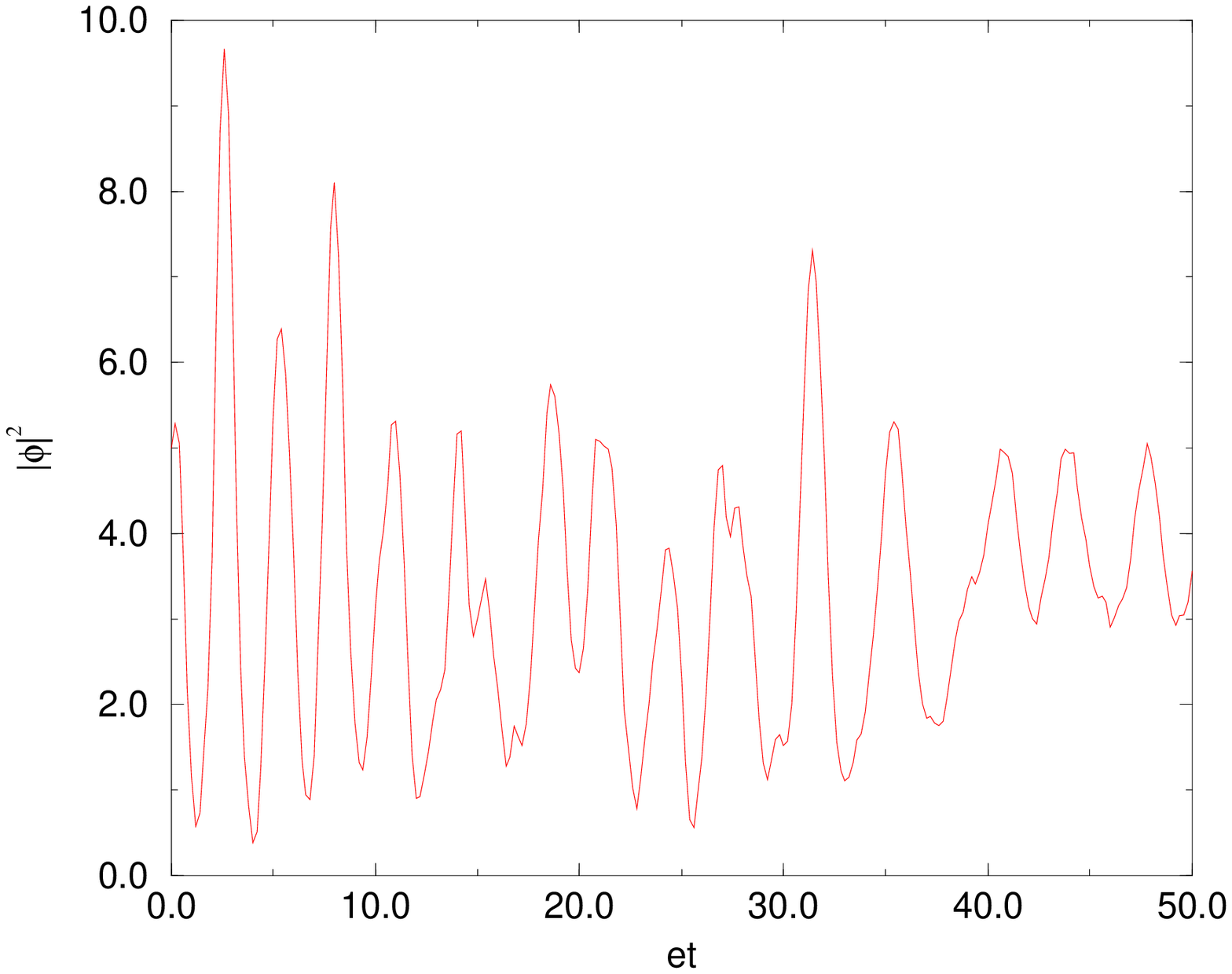,width=8.5cm}
\vspace{-0.4cm}
\caption{\narrowtext
Order parameter $|\phi|^2$ versus $et$.}
\label{fig3a}
\end{figure}

Interesting observables are the Chern-Simons number $C$ and the 
anomalous axial charge $Q_5$. According to the anomaly equation, these should 
be equal during time evolution.
$C$ is a 'simple' observable in the sense that it is just the sum of 
$A_1$ over all lattice points.
On the contrary, $Q_5$ is obtained by taking the sum over inner 
products between  all modes functions, see (\ref{eqq5}). Results
are given in Fig.~\ref{fig3}. 
It shows that $Q_5$ follows $C$ at 
early times, in agreement with the anomaly equation. For later times they 
start to deviate. This is an artefact of the lattice discretization, 
which is reduced by taking smaller lattice spacing.
In Fig.~\ref{fig3a} we show the Higgs order parameter.

\vspace{-0.4cm}

\section{Conclusions and outlook}

We considered the real-time dynamics of a coupled system of classical 
Bose fields and a quantized fermion field, represented by generalized 
mode functions, on a lattice in space and time.  
Because of the presence of $2N$ spinor mode functions that are space 
dependent, the method is, especially in $3+1$ dimensions, numerically 
demanding.

The $1+1$ dimensional model is an excellent toy model to test the 
possibility of numerical calculations in a non-homogeneous, non-equilibrium 
context, including anomalous fermion production. 
A more detailed report will appear elsewhere \cite{AaSm98}.

\acknowledgments
This work is supported by FOM.

\end{multicols}

\begin{references}
\bibitem{GrRu88}
	D.Yu.~Grigoriev and V.A.~Rubakov,
        Nucl.\ Phys.\ B299 (1988) 67.
\bibitem{abh}
	D.Yu.~Grigoriev, V.A.~Rubakov and M.E.~Shaposhnikov, 
        Nucl.\ Phys.\ B326 (1989) 737; 
	Ph.~de Forcrand,  A.~Krasnitz and R.~Potting, 
        Phys.\ Rev.\ D50 (1994) 6054;
	J.~Smit and W.H.~Tang, 
        Nucl.\ Phys.\ B (Proc.\ Suppl.) 34 (1994) 616; ibid. 42 (1995) 590. 
\bibitem{TaSm98}  
	W.H.~Tang and J.~Smit, hep-lat/9805001.
\bibitem{su2}
	J.~Ambj\o rn, T.~Askgaard, H.~Porter and M.E.~Shaposhnikov, 
	Phys. Lett. B244 (1990) 497;  Nucl.\ Phys.\ B353 (1991) 346;
	J.~Ambj\o rn and A.~Krasnitz,  Phys. Lett. B362 (1995) 97;  
	Nucl.\ Phys.\ B506 (1997) 387;
	W.~H.~Tang and J.~Smit, 
	Nucl.\ Phys.\ B482 (1996) 265; ibid. B51 (1998) 401; 
        G.D.~Moore and N.~Turok,
	Phys.\ Rev.\ D55 (1997) 6538; ibid. D56 (1997) 6533. 
\bibitem{AaSm97}
	G.~Aarts and J.~Smit, 
	Phys.\ Lett.\ B393 (1997) 395; 
  	Nucl.\ Phys.\ B511 (1998) 451.
\bibitem{classhtl1}
	  D.~B\"odeker, L.~McLerran and A.~Smilga,
          Phys.\ Rev.\ D52 (1995) 4675;
	  P.~Arnold, D.~Son, L.G.~Yaffe, Phys.\ Rev.\ D55 (1997) 6264; 
	  P.~Arnold, Phys.\ Rev.\ D55 (1997) 7781.
\bibitem{classhtl2}
   	  C.R.~Hu and B.~M\"uller, Phys.\ Lett.\ B409 (1997) 377; 	
	  G.D.~Moore, C.R.~Hu and B.~M\"uller, 
	  Phys.\ Rev.\ D58 (1998) 45001;
          E.~Iancu, hep-ph/9710543.
\bibitem{CoMo1}
	F.~Cooper and E.~Mottola, 
	Phys.\ Rev.\ D36 (1987) 3114.
\bibitem{CoMo2}
	F.~Cooper, S.~Habib, Y.~Kluger, E.~Mottola, J.P. Paz and 
	P.R. Anderson, Phys.\ Rev.\ D50 (1994) 2848.
\bibitem{largeN}
	Y.~Kluger, J.M.~Eisenberg, B.~Svetitsky, F.~Cooper and E.~Mottola, 
	Phys.\ Rev.\ D45 (1992) 4659;
	F.~Cooper, S.~Habib, Y.~Kluger and E.~Mottola, 
	Phys.\ Rev.\ D55 (1997) 6471;
	D.~Boyanovsky, H.J.~de~Vega, R.~Holman and J.F.J~Salgado, 
	Phys.\ Rev.\ D54 (1996) 7570;
	D.~Boyanovsky, D.~Cormier, H.J.~de~Vega, R.~Holman, A.~Singh and 
	M.~Srednicki, Phys.\ Rev.\ D56 (1997) 1939.
\bibitem{baacke}
  	J.~Baacke, K.~Heitmann and C.~Patzold, 
	Phys.\ Rev.\ D55 (1997) 7815; ibid. D57 (1998) 6406; 
	hep-ph/9806205.
\bibitem{AaSm98}
	G.~Aarts and J.~Smit, 
	{\em Real-time dynamics with fermions on a lattice}, to appear.
\end{references}
\end{document}